\documentclass{article}

\usepackage{arxiv}

\usepackage[utf8]{inputenc} 
\usepackage[T1]{fontenc}    
\usepackage{hyperref}       
\usepackage{url}            
\usepackage{booktabs}       
\usepackage{amsfonts}       
\usepackage{nicefrac}       
\usepackage{microtype}      
\usepackage{lipsum}		
\usepackage{graphicx}
\usepackage[numbers,sort&compress]{natbib}
\usepackage{doi}
\usepackage{caption}
\usepackage{subcaption}

\title{Automation of radiation treatment planning for rectal cancer}
\author{ \href{https://orcid.org/0000-0002-7683-5093}{\includegraphics[scale=0.06]{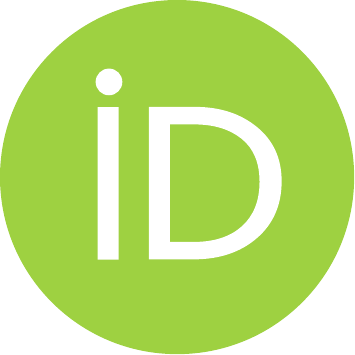}\hspace{1mm}Kai Huang}\\
	The University of Texas MD Anderson Cancer Center UTHealth\\
	Graduate School of Biomedical Sciences\\
	Houston, TX 77030 \\
	\texttt{khuang5@mdanderson.org} \\
	\And
	{Prajnan Das} \\
	Department of Radiation Oncology\\
	The University of Texas MD Anderson Cancer Center\\
	Houston, TX 77030 \\
	\texttt{prajdas@mdanderson.org} \\
	\And
	{Adenike M. Olanrewaju}\\
	Department of Radiation Physics\\
	The University of Texas MD Anderson Cancer Center\\
	Houston, TX 77030 \\
	\texttt{aolanrew@mdanderson.org}\\
	\And
	{Carlos Cardenas}\\
	Department of Radiation Oncology\\
	The University of Alabama at Birmingham\\
	Birmingham, AL 35233 \\
	\texttt{cecardenas@uabmc.edu}\\
	\And
	{David Fuentes}\\
	Department of Imaging Physics\\
	The University of Texas MD Anderson Cancer Center\\
	Houston, TX 77030 \\
	\texttt{dtfuentes@mdanderson.org}\\
	\And
	{Lifei Zhang, Donald Hancock}\\
	Department of Radiation Physics\\
	The University of Texas MD Anderson Cancer Center\\
	Houston, TX 77030 \\
	\texttt{{lifzhang, dhancock}@mdanderson.org}\\
	\And
	{Hannah Simonds}\\
	Department of Radiation Oncology\\
	Tygerberg Hospital/University of Stellenbosch\\
	Stellenbosch, South Africa \\
	\texttt{hsimonds@sun.ac.za}\\
	\And
	{Dong Joo Rhee, Sam Beddar, Tina Marie Briere, Laurence Court}\\
	Department of Radiation Physics\\
	The University of Texas MD Anderson Cancer Center\\
	Houston, TX 77030 \\
	\texttt{{drhee1, abeddar, tmbriere, lecourt}@mdanderson.org}\\
}

\renewcommand{\headeright}{}
\renewcommand{\undertitle}{}

\hypersetup{
pdftitle={Automation of radiation treatment planning for rectal cancer},
pdfauthor={Kai Huang},
pdfkeywords={First keyword, Second keyword, More},
}

\begin{document}
\maketitle

\begin{abstract}
To develop an automated workflow for rectal cancer three-dimensional conformal radiotherapy (3DCRT) treatment planning that combines deep learning (DL) aperture predictions and forward-planning algorithms.
We designed an algorithm to automate the clinical workflow for 3DCRT planning with field aperture creations and field-in-field (FIF) planning. DL models (DeepLabV3+ architecture) were trained, validated, and tested on 555 patients to automatically generate aperture shapes for primary (posterior–anterior [PA] and opposed laterals) and boost fields. Network inputs were digitally reconstructed radiographs, gross tumor volume (GTV), and nodal GTV. A physician scored each aperture for 20 patients on a 5-point scale (>3 is acceptable). A planning algorithm was then developed to create a homogeneous dose using a combination of wedges and subfields. The algorithm iteratively identifies a hotspot volume, creates a subfield, calculates dose, and optimizes beam weight all without user intervention. The algorithm was tested on 20 patients using clinical apertures with varying wedge angles and definitions of hotspots, and the resulting plans were scored by a physician. The end-to-end workflow was tested and scored by a physician on another 39 patients.
The predicted apertures had Dice scores of 0.95, 0.94, and 0.90 for PA, laterals,and boost fields, respectively.Overall,100\%,95\%,and 87.5\% of the PA, laterals, and boost apertures were scored as clinically acceptable, respectively. At least one auto-plan was clinically acceptable for all patients. Wedged and non-wedged plans were clinically acceptable for 85\% and 50\% of patients, respectively. The hotspot dose percentage was reduced from 121\% ($\sigma =$ 14\%) to 109\% ($\sigma$ = 5\%) of prescription dose for all plans. The integrated end-to-end workflow of automatically generated apertures and optimized FIF planning gave clinically acceptable plans for 38/39 (97\%) of patients.
We have successfully automated the clinical workflow for generating radiotherapy plans for rectal cancer for our institution. 

\end{abstract}


\section{Introduction}
Colorectal cancer is the third most common cancer globally and ranks second for mortality, with more than 1.9 million cases diagnosed each year. Of those, more than 700 000 are rectal cancers.\cite{sung_global_2021,hong_radiation_2017} Rectal cancer is typically treated with a combination of chemoradiotherapy and surgery. The most common radiotherapy technique is three-dimensional conformal radiotherapy (3DCRT).\cite{hong_radiation_2017, wo_radiation_2020} Each radiotherapy treatment must be individualized and carefully planned to precisely deliver radiation to tumor while minimizing toxicity to surrounding healthy tissues. This requires homogenous dose distribution with sufficient dose coverage of the target volume. The treatment planning process can be considered in two steps: determination of treatment field apertures, then the use of accessories or subfields to give a homogeneous dose distribution.\par
For rectum radiotherapy, beams are placed in either 3- or 4-field geometry, with or without an anterior-posterior (AP) beam.\cite{hong_radiation_2017, gearhart_early_2010,minsky_pelvic_1993} The designs of the radiation treatment fields for rectal cancers are determined by a physician and can vary across practices.\cite{hong_radiation_2017,minsky_pelvic_1993, nakamura_strategy_2019, yu_dosimetric_2013} Regardless of these differences, the boundaries of treatment fields depend on the location of the gross tumor volume (GTV), involved lymph nodes (GTVn), and bony landmarks. Researchers have developed solutions for defining patient-specific apertures for other cancers. For example, Kisling et al. developed multi-atlas and deep learning (DL) solutions for cervical cancer, with the field shape based on bony landmarks.\cite{kisling_fully_2019, kisling_automatic_2020} Cardenas et al. have compared different convolutional neural networks to automatically define patient-specific apertures for cervical cancer where the boundaries of treatment fields depend solely on bony landmarks.\cite{cardenas_comparison_2018} No previous studies have automated the manual placement of apertures for rectal cancers.\par
Once the field apertures have been determined, beam weights are adjusted through trial and error. Owing to asymmetries in human anatomy in the sagittal direction, undesirable high-dose clouds (hotspots) usually occur in the posterior region of the body. To minimize hotspots and improve dose homogeneity in the target volume, wedges or field-in-field (FIF) techniques can be used. FIF is a treatment planning technique for hotspot reduction and dose conformity that is widely used in breast,cervix,and rectum radiotherapy planning, among others.\cite{prabhakar_field--field_2009, onal_dosimetric_2012, yavas_dosimetric_2013, mayo_forward-planned_2004} FIF technique can reduce hotspots while maintaining adequate dose coverage to the target volume.\cite{yavas_dosimetric_2013} Compared with wedges, FIF provides more nuanced hotspot reduction and can achieve an equivalent or better dose distribution.\cite{prabhakar_field--field_2009,onal_dosimetric_2012,prabhakar_can_2008} FIF can be preferable in clinical practice to a dynamic wedge because of its ability to operate in two dimensions instead of one.\cite{torre_comparative_2004,softa_p293_2018} However, the process of manually adjusting beam weights and adding subfields is time-consuming and highly dependent on planner expertise. Automated solutions to add subfields exist for breast plans.\cite{kim_automated_2018, kisling_automated_2019} To our knowledge, our study is the first to describe an automated solution with beam weight optimization to add wedges and FIF to rectal plans.The goal of this study is to design and implement an automatic end-to-end solution for 3DCRT planning of rectal cancer and produce clinically acceptable plans reviewed by a gastrointestinal (GI) radiation oncologist.

\section{Methods}\label{sec:Methods}
We developed an automatic approach to generate field apertures based on CT and clinical GTV and GTVn, replicating the manual drawing process. We then developed a method to perform 3DCRT planning automatically, including primary and boost fields, beam weight optimization, wedges, and FIF.We assessed the performance of the field aperture generation and the FIF planning algorithm independently, and as a whole.

\subsection{Field aperture}
\begin{figure}
	\centering
	\includegraphics[scale=.5]{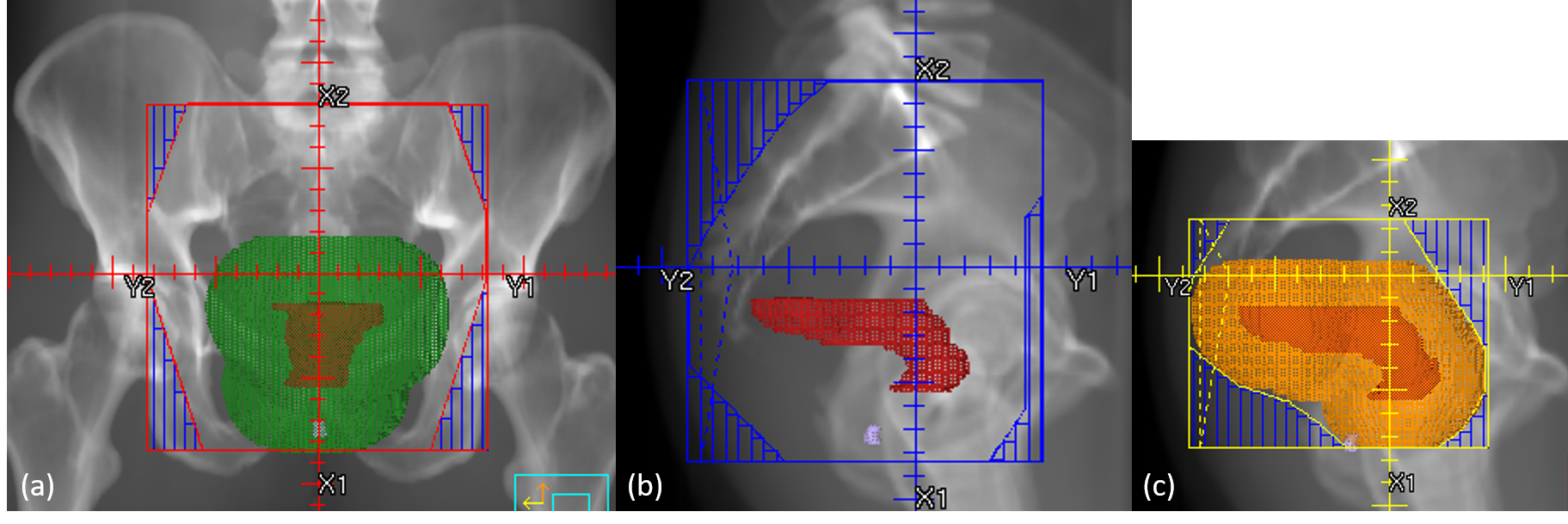}
	\caption{The figure shows the clinical guidelines for field aperture placement, including primary beams and boost beams. The primary beams have one posterior–anterior (PA) beam and two opposed lateral beams. The boost beams have two opposed lateral beams. The opposed lateral beams are mirror images of each other: (a) PA beam, (b) primary right lateral beam, and (c) right boost beam. The red structure is gross tumor volume (GTV), the green structure is 3-cm uniform expansion of GTV, and the orange structure is 2-cm uniform expansion of GTV.}
	\label{fig:fig1}
\end{figure}

Our clinical practice uses bony anatomy and the locations of GTV and GTVn to determine the field apertures. The information about bony landmarks is derived from digitally reconstructed radiographs (DRRs) at each beam angle. Figure \ref{fig:fig1} illustrates how field apertures are determined for each beam angle in our clinic. Specifically, primary beams include a posterior–anterior (PA) beam and two opposed lateral beams. For the PA beam (Figure \ref{fig:fig1}a), its superior border is set at either the L5/S1 intervertebral body intersection or 2-cm superior to GTV or GTVn,whichever is more superior.The inferior border is set at either the bottom of the obturator foramen or 3-cm inferior to the GTV, whichever is more inferior. The lateral borders are set at $\sim$2 cm from the pelvic brim. Once the borders are decided for the PA beam, the four corners of the field aperture are blocked off to prevent excessive dose to the sacroiliac joint and femoral heads. The primary lateral beams include left (LT) and right (RT) laterals and are usually mirror images of each other (Figure \ref{fig:fig1}b). The superior and inferior borders of lateral beams match with those of the PA beam. The posterior border of the lateral beams covers the entire sacrum and extends 1-cm posterior to the sacrum. The anterior border of the lateral beams extends to 3-cm anterior to the sacral promontory, or 2-cm anterior to GTV or GTVn, whichever is more anterior. The boost beams (Figure \ref{fig:fig1}c) are also opposed lateral beams, including both left (BstLT) and right (BstRT) laterals. The boost fields should include the GTV,GTVn,and adjacent pelvic and presacral regions to GTV and GTVn.

\begin{table}
	\caption{The table shows the number of patients, training, validation and testing split for each field. Abbreviation: PA, posterior-anterior.}
	\centering
	\begin{tabular}{lllll}
		\toprule
		Field	& Train	& Validation	&Test	&Total\\
		\midrule
		PA	&404&	60&	60&	524\\
        Laterals	&403	&60	&60&	523\\
        Boost	&435&	60&	60&	555\\
		\bottomrule
	\end{tabular}
	\label{tab:table1}
\end{table}
We mimicked this clinical decision-making process by using DeepLabV3+\cite{ferrari_encoder-decoder_2018}, a convolutional neural network, with DRR and expanded clinical GTV and GTVn as inputs to predict the field apertures for each field. We used 3DCRT plans collected retrospectively from a total of 555 rectal patients. The plans were curated such that only the ones following the aforementioned guidelines for determining field apertures were used in the training process. The number of patients and the training, validation, and testing split for each field are shown in Table \ref{tab:table1}.\par

\begin{figure}
\centering
  \includegraphics[scale=.6]{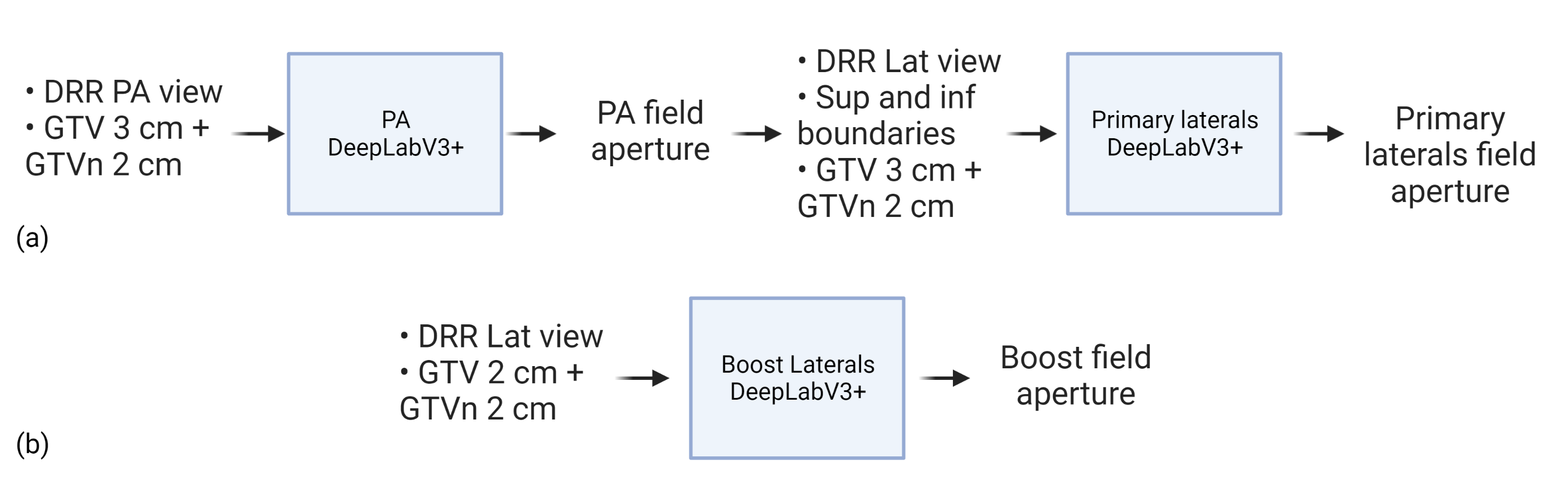}
  \caption{The figure shows the relationship between each model, and the inputs and outputs of each model. Either the mask of 3- or 2-cm-expanded gross tumor volume (GTV) combined with 2-cm-expanded GTV involved lymph nodes (GTVn) were used as inputs to the DeepLabV3+ architecture.\cite{ferrari_encoder-decoder_2018} Part (a) shows the inputs and outputs for the posterior–anterior (PA) and primary lateral (Lat) models. Part (b) show the inputs and outputs for the boost Lat model. DRR, digitally reconstructed radiograph; Sup, superior; Inf, inferior}
  \label{fig:fig2}
\end{figure}

We trained one model for the PA field, one model for primary laterals, and one model for boost laterals. Optimal learning rates were determined by training multiple models with varying learning rates. To match the superior and inferior borders of the primary fields, the primary models were sequential, meaning that the superior and inferior borders predicted by the model for the PA field were used as an additional input in the form of binary masks for the lateral model. The relationship between models and the inputs for each are illustrated in Figure \ref{fig:fig2}.
For model inputs, we used an in-house DRR generator to generate both DRR and masks for expanded GTV and GTVn. Uniform expansions of 3 and 2 cm were applied to the primary GTV and GTVn, respectively. These masks were then combined and used as an input for the primary models. For the boost model, 2-cm expanded GTV and GTVn were combined as a single mask and inputted into the architecture. For primary lateral and boost fields, the models predicted left and right fields separately.\par
\begin{table}
	\caption{The table shows the 5-point scale of evaluating the quality of predicted field apertures and plans}
	\centering
	\begin{tabular}{lll}
		\toprule
Score&	Acceptability&	Description\\
\midrule
5&	Acceptable, use as-is	&Clinically acceptable,\\
&&could be used for treatment without change\\
4&	Acceptable, minor edits that are not necessary&	Stylistic differences, but not clinically important; \\
&&the current contours/plans are acceptable\\
3&	Unacceptable, minor edits that are necessary&	Edits that are clinically important;\\&& but it is more efficient to edit the automatically \\&&generated contours/plans than to start from scratch\\
2&	Unacceptable, major edits&	Edits that are required to ensure appropriate treatment\\&& and sufficiently significant that the user would prefer\\&& to start from scratch\\
1&	Unacceptable, unusable	&Automatically generated contours/plans are so bad \\&&that they are unusable\\&& (i.e., wrong body area, outside confines of body, etc.)\\
		\bottomrule
	\end{tabular}
	\label{tab:table2}
\end{table}

The architecture was trained using Dice similarity coefficient loss, which is a widely used loss function for segmentation tasks and provides stable training for our task.\cite{jadon_survey_2020} Dice similarity coefficient (Dice), Hausdorff distance (HD), 95th percentile HD (HD95), and mean surface distances (MSD) were calculated to quantitatively evaluate the results of the models.\cite{taha_metrics_2015} No post-processing was applied to the aperture shapes before multi-leaf collimators (MLC) fitting. Each aperture was evaluated and scored by a GI radiation oncologist on a 5-point scale summarized in Table \ref{tab:table2}.

\subsection{Field-in-field}
Following a conventional clinical workflow, after field apertures are determined, the beam weights for the primary apertures are adjusted, and wedges, FIF, or both are added to reduce hotspots. Typically, an initial dose calculation is performed with primary fields. Then the hotspots, projected into a beam’s eye view of a treatment field,are blocked by adding subfields.The definition of hotspots as a percentage of the prescription dose (Rx) depends on the treatment sites and the clinical practices. This is a value beyond which dose clouds would be considered to be hotspots and unacceptable if the sizes are too large. After this step, the beam weights for both primary and subfields need to be readjusted with the goal of increasing homogeneity of planning target volume (PTV) while decreasing the hotspot volumes. This entire process is iterative.\par
Here, we describe our approach to automatic FIF optimization to improve the homogeneity of the dose distribution. We will describe the FIF algorithm then detail how the program was configured for testing. Adding boost fields follows the same algorithmic steps as the primary fields; therefore, the experiment focused on testing and assessing the primary fields.

\subsection{Algorithm}
\subsubsection{Overview}

\begin{figure}
\centering
  \includegraphics[scale=.49]{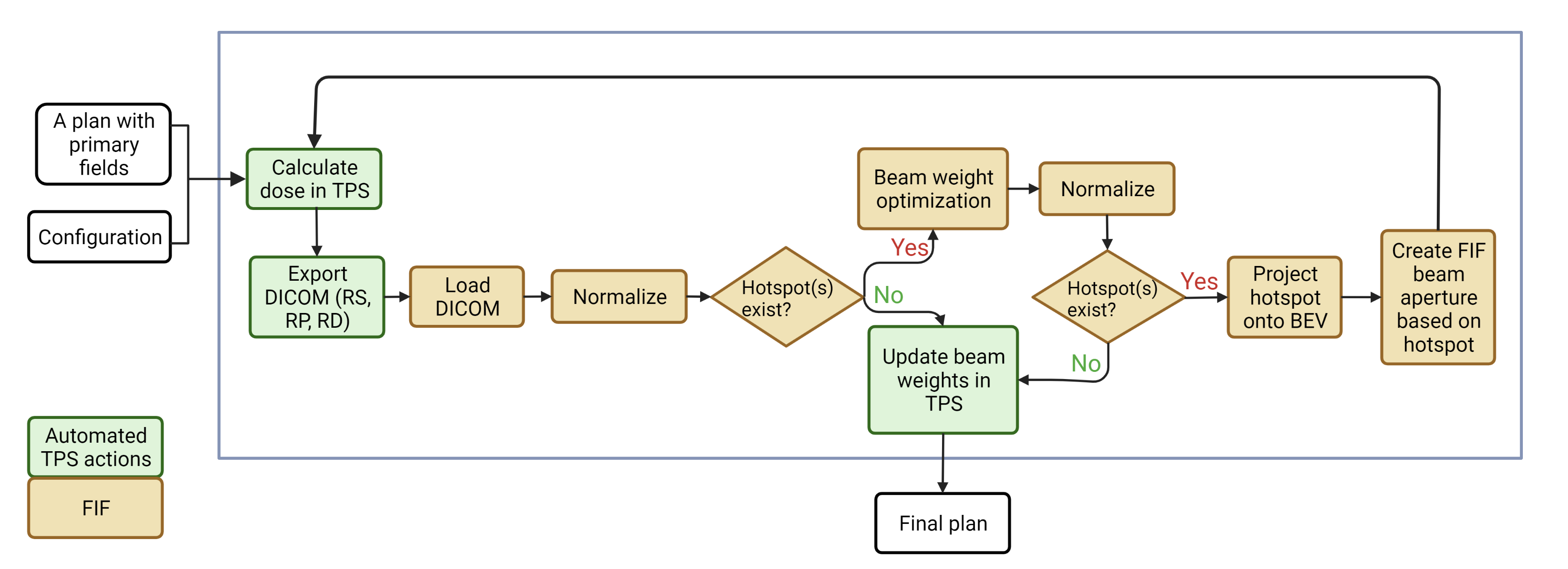}
  \caption{The figure shows the flowchart of the field-in-field (FIF) algorithm. TPS, treatment planning system; BEV, beam’s eye view; DICOM, Digital Imaging and Communications in Medicine; RS, DICOM RT Structure file; RP, DICOM RT Plan file; RD, DICOM RT Dose file}
  \label{fig:fig3}
\end{figure}

An overview of the FIF algorithm workflow can be seen in Figure \ref{fig:fig3}. The design of the algorithm took a forward-planning approach to mimic the clinical process. The goal of the program is to minimize hotspots in the plan and achieve a target dose percentage. The program communicates with treatment planning systems (TPSs) via Digital Imaging and Communications in Medicine (DICOM) and their respective Python Application Programming Interfaces. The program is compatible with both Eclipse and RayStation TPSs. The FIF algorithm was tested in RayStation version 10B and Eclipse version 15.5. The FIF algorithm was written in CPython version 3.6. The SimpleITK package was used to read DICOM dose files and perform image transformation and manipulations.\par

The algorithm loads a 3DCRT plan in the TPS with predefined apertures for primary fields. The locations of jaws and MLC of primary fields are not modified by the algorithm. The number of primary fields is not restricted, which makes the program versatile for use in rectal planning with either 3- or 4-field geometries.\par
The actions of the program are dictated by a list of configurable variables. Essential configurable variables include the specific plan to load, prescriptions,maximum allowed iteration, subfield minimum and maximum monitor unit (MU), target dose percentage, whether to add wedges or to normalize by volume or point (Equation \ref{eq:normalization}), and the optimization solver. This allows the algorithm to be easily customized for various planning requirements and restrictions for different body sites and clinics.Upon normalization,the dose for each voxel would be updated as follows:
\begin{equation}\label{eq:normalization}
    D_i \rightarrow  Rx \times \frac{D_i}{D_p}
\end{equation}
where Rx represents the prescription dose in Gy, $D_i$ represents the dose in Gy for each voxel, and $D_p$ (Gy) is either the dose at the specified point of a region of interest (ROI) if using normalization by point or the dose at a specific percentage volume of an ROI, if using normalization by volume.\par
As shown in Figure \ref{fig:fig3}, processes, such as dose calculation, importing and exporting of DICOM files, beam weight updates, adding of wedges, and MLC fitting, are automatically performed within the TPS by the FIF program. The exported DICOM files include RT Structure, RT Plan, and RT Dose files for each beam. These steps are performed using the scripting interface provided by the TPS and are an integral part of the FIF program.The FIF algorithm is modularized so that changing the TPS requires only a change of setting in the configuration file.\par
After exporting the relevant DICOM files, the FIF algorithm loads the DICOM files and checks if a hotspot exists for the given beams and beam weights. As the location and size of the hotspots are highly dependent on the normalization of the plan, the program always renormalizes before a hotspot existence check is performed. If there is no hotspot, then the program terminates. If there is at least one hotspot, the program performs beam weight optimization to minimize the hotspot first then creates a subfield aperture based on the minimized hotspot location and size.The information on beam weight optimization,hotspot determination, and how the subfield aperture is created are detailed in the following sections. Once the hotspot to block is determined and projected into the beam’s eye view, its 2D shape is imported into the TPS for MLC fitting and dose calculation. The entire process iterates until there is no hotspot left or the maximally allowed number of subfields has been added, as shown in Figure \ref{fig:fig3}.\par
The algorithm terminates under three conditions: (1) the desired percentage hotspot dose has been reached, (2) the maximum number of subfields has been added, or (3) adding another subfield will increase the final percentage hotspot dose. In cases where the desirable percentage hotspot dose is too difficult to achieve, the program could terminate prematurely based on the last condition. In such cases, percentage hotspot dose can be further reduced by relaxing the goal, that is, increasing the target dose percentage. Users can configure the algorithm to relax the goal and continue optimizing. By default, the target dose percentage is increased by 1\%. 

\subsubsection{Beam weight optimization}
Beam weight optimization uses quadratic optimization as described by the following equation.
\begin{equation}
f = \sum^{M}_{m} \omega_m \sum^{I}_{i}(\sum^{J}_{j} \alpha_j B_{ij}-D^{Rx}_{m} )^2  \quad \quad \frac{min MU}{total MU}\leq \alpha_j\leq \frac{max MU}{total MU}
\end{equation}
where m is the $m^{th}$ ROI, $\omega_m$ is the weight given to each ROI, i represents the $i^{th}$ voxel in the ROI, j represents the $j^{th}$ number of beam, $\alpha_j$ is the beam weight for the $j^{th}$ beam, $B_{ij}$ is the beam dose (Gy) for a specific voxel, and $D_m^{Rx}$ (Gy) is the specified dose for the $m^{th}$ ROI, which is a percentage of the total Rx for the plan. Each beam weight, $\alpha_j$, has boundaries defined by the minimum and maximum MUs according to the given configuration. The function includes a step function term. For target volumes, only the voxels lower than Rx are penalized. For organs at risk (OARs), only the voxels higher than the specified tolerance dose are penalized. Voxels overlapped by primary beams are penalized if the values exceed the definition of hotspot. ROIs can be constructed and modified for different treatment sites. Specific OARs and targets used in this experiment are described in the experimental setup section.\par
As shown in Figure \ref{fig:fig3}, for each iteration, the optimal beam weights are calculated if hotspots still exist. To solve the optimization problems, different solvers in the SciPy package can be specified in the configuration file and used.By default,none of the previously created field weights is fixed, and all of the beam weights are subject to change after running through the beam weight optimization. 

\subsubsection{Determining hotspots}
Hotspots were found by thresholding the plan dose by the configured hotspot percentage at each iteration. A hotspot is defined by default to be 107\% of Rx with at least 8 $cm^3$ in volume fully connected based on clinical practices in our institution. At each iteration, subfields can be made based on a fixed percentage of a hotspot,for example,reducing a 107\% hotspot iteratively till it is nonexistent, or the percentage of the hotspot can decrease at each iteration at a fixed decrement, for example, subsequently creating subfields based on 115\%, 110\%, and 105\% Rx. Users can configure the algorithm to use either method. Therefore, the definition of hotspots may be fixed or changed at each iteration, and the identified hotspots would be used to create the shape of a subfield for the iteration.\par
As hotspots are identified through dose thresholding, the 3D hotspot volume mask needs to be transformed into the beam’s eye view and projected into 2D for the field on which the subfield is based. After the transformation, the distal portion of the 3D hotspot mask with respect to the isocenter is truncated. The reason is to minimize the possibility of causing coldspots in the target volume when the hotspots are proportionally larger on the distal side of the isocenter.At this stage, the users can configure the algorithm such that only the largest hotspot will be retained. This option is available because sometimes the existence of excessive air pockets can break up a single hotspot into distant smaller volumes. In such scenarios, covering the largest hotspot volume is preferable to covering all the volumes because cold spots may appear, causing the program to terminate, as the given normalization would no longer be achievable. After the truncation,the transformed hotspot is projected onto the iso-centric plane using ray tracing projection. The resulting 2D hotspot mask is used to create a single aperture shape for an MLC to conform to.\par
The aperture shape with x and y coordinates is exported into TPS for calculation of MLC positions. The single shape is constructed by traversing over all the disconnected hotspots in the 2D mask and using the closest MLC bank to cover the hotspot object. After the shape is exported, a subfield is added into the plan in a TPS, and the coordinates of the shape are imported into the newly added subfield for MLC fitting.\par
The percentage for target dose may be different from the hotspot definition. Once the target dose percentage is reached, the program terminates. The target dose is the hotspot dose the program strives for, meaning that the maximum value of dose in at least 8 $cm^3$ fully connected volume should be less or equal to the percentage for target dose. For example, a target dose of 107\% of Rx means that the program will terminate once the maximum hotspot dose is less or equal to 107\% of Rx. By default, the target dose percentage is set to be 107\% of Rx. The hotspot, target dose percentage, and the size of the minimum hotspot are configurable and can be changed to suit different clinical requirements.

\subsection{Experimental setup}\label{sec:sec2p4}
\begin{figure}
\centering
  \includegraphics[scale=1.2]{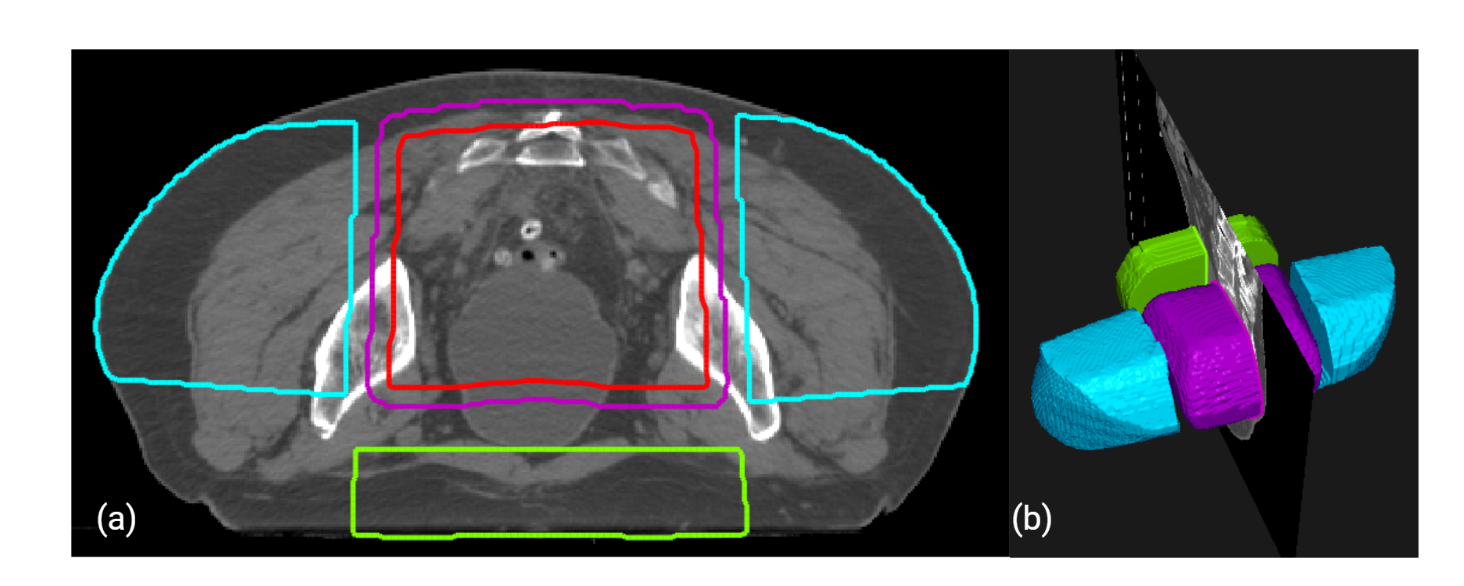}
  \caption{The figure shows an example of the constructed volumes for the experimental setup. The purple volume is the region of high dose, the red volume is pseudo-region of interest (ROI), the blue volumes are LTRT, and the green volume shows APPA. Part (a) shows a transverse slice, and (b) shows a three-dimensional view of the volumes.}
  \label{fig:fig4}
\end{figure}
In this study, we independently evaluated the performance of the FIF program by retrospectively testing the program on 20 patients using clinical apertures with 3-field geometry without boost fields. The following structures were constructed by ROI algebra. The region of high dose (RHD) is the volume overlap of primary beams. RHD is used to penalize hotspots. PTV is not contoured before the FIF algorithm is run; therefore, the program constructs a pseudo-ROI (pROI) by uniformly eroding the RHD by 1 cm.To achieve dose objectives for regions, including femur and bowel space without having to create accurate contours,two structures named LTRT and APPA were introduced. LTRT structures are the two regions on the lateral sides of the PTV, formed by the overlap of the lateral beams minus the 1-cm expansion of RHD.Finally,APPA is the region on the anterior side of the PTV, formed by the overlap of the AP and PA beams minus the 1-cm expansion of RHD. As there is no AP beam in the 3-field geometry, only the PA beam is used to construct APPA. Figure \ref{fig:fig4} shows an example of the aforementioned volumes.\par
To ensure fair comparison, all the plans were initialized with 2:1:1 beam weighting ratios for PA, LT lateral, and RT lateral beams. After beam weight initialization and initial beam weight optimization, all plans were normalized such that the prescribed dose covered 99\% of the pROI. For beam weight optimization, an L-BFGS-B solver, which is in the family of quasi-Newton optimization methods,was used. For hotspot projection, only the largest hotspot was used to construct an MLC-fitting shape. The algorithm added one subfield per iteration to only the LT and RT sides in an alternating manner. A TrueBeam machine (Varian Medical Systems) with standard 120 millennium MLC and 15-MV photon beams were used for constructing plan and dose calculation. The maximum number of subfields added was set to be 6. For every iteration, only one subfield is added. The experiments were conducted in RayStation TPS using collapsed cone dose calculation algorithm at each iteration. The percentage of target dose was defined to be 107\% of Rx,which was 45 Gy in 25 fractions for primary beams.\par
To evaluate the effect of using different wedges in the plan, we planned with (1) a 45-degree wedge, (2) a 60-degree wedge, and (3) no wedge, other things being equal, for each patient.Wedges were only added to both primary lateral fields, and the same angle was used for each side. \par
To evaluate the effect of defining hotspot percentage differently, we compared plans with (1) 107\% of Rx defined as a hotspot and (2) 106\% of Rx defined as a hotspot, whereas the target dose percentage was kept at 107\% and wedges at 45 degrees, other things being equal. The hotspot definition was kept the same throughout iterations. This means that for plans having 106\% of Rx defined as a hotspot, 106\% of Rx was used to perform dose thresholding and create subfields for all iterations.\par
In total, four plans were created for each patient. The final hotspot percentage and maximum point dose percentage were calculated and compared for each plan. The time it took for each plan to complete was also recorded. All summary measures are shown as mean and standard deviation unless otherwise indicated. All plans were evaluated and scored by a radiation oncologist on a 5-point scale (Table \ref{tab:table2}) based on the final dose of the primary beam set.

\subsection{End-to-end workflow}
Once the automatic aperture generation and FIF optimizations had been separately developed and tested, we performed a complete end-to-end test of the system. To do this,we used retrospective data from another 39 patients with clinical GTV and GTVn predefined. The field apertures predicted by DL models were used as inputs to TPS, the FIF program made one plan for each patient using 60-degree wedges, and hotspot percentage was defined as 107\% of Rx, all other things being equal to configurations described in Section \ref{sec:sec2p4}. The beam weight optimization algorithm determined the weightings of all the beams.\par
Different from the evaluation of the FIF algorithm alone, the end-to-end evaluation used the FIF algorithm to add boost fields with apertures predicted by DL models. Following our clinical guidelines,a 15-degree wedge was added to each of the lateral boost fields, and no additional subfields were added. The Rx was set to be 5.4 Gy in three fractions for boost beams.The normalization was set to be the 100\% of Rx isodose line covering 98\% of the boost volume,which was constructed by the overlap of volume from boost apertures eroded by 1 cm and pROI. The plans were evaluated and scored by a radiation oncologist on a 5-point scale (Table \ref{tab:table2}) based on the final plan dose.

\section{Results}
\subsection{Field aperture}
\begin{table}
	\caption{The table shows the average metrics results for field aperture prediction. Abbreviations: Dice, Dice coefficient; HD, Hausdorff distance; HD95, 95th percentile of Hausdorff distance; MSD, mean surface distance; PA, posterior–anterior.}
	\centering
	\begin{tabular}{lllll}
		\toprule
Model	&Dice	&HD (mm)	&MSD (mm)	&HD95 (mm)\\
		\midrule
	PA	&0.95&	13.8&	3.7	&8.6\\
Laterals&	0.94&	15.5&	4.6&	11.8\\
Boost	&0.90	&18.7	&6.0	&15.4\\
		\bottomrule
	\end{tabular}
	\label{tab:table3}
\end{table}
\begin{table}
	\caption{The table shows the results of physician reviews and scoring for each aperture field. Abbreviations: BstLT, left lateral boost; BstRT, right lateral boost; LT, primary left
lateral; PA, posterior–anterior;RT, primary right lateral.}
	\centering
\begin{tabular}{llllllll}
\toprule
         & \multicolumn{5}{l}{Counts per score} & \multicolumn{2}{l}{Acceptability percentage} \\
\midrule
Aperture & 5      & 4     & 3     & 2    & 1    & Acceptable (\%)           & Unacceptable (\%)          \\
\midrule
PA       & 16     & 4     & 0     & 0    & 0    & 100                & 0                   \\
LT       & 11     & 8     & 1     & 0    & 0    & 95                 & 5                   \\
RT       & 11     & 8     & 1     & 0    & 0    & 95                 & 5                   \\
BstLT    & 14     & 3     & 3     & 0    & 0    & 85                 & 15                  \\
BstRT    & 13     & 5     & 2     & 0    & 0    & 90                 & 10\\
\bottomrule
\end{tabular}
	\label{tab:table4}
\end{table}
PA, laterals, and boost models used 97, 60, and 64 epochs for model training, respectively. The results of overlap and distance metrics for each model are shown in Table \ref{tab:table3}. The predicted apertures had Dice average scores of 0.95, 0.94, and 0.90 for PA, primary lateral, and boost fields, respectively (Table \ref{tab:table3}). The predicted apertures had average HD, MSD, and HD95 of below 1.6, 0.5, and 1.2 cm for primary fields, respectively. The boost fields had average HD,MSD,and HD95 below 1.9, 0.6, and 1.5 cm, respectively. \par
As shown in Table \ref{tab:table4}, 100\% of PA apertures were clinically acceptable. None of the apertures were scored at 2 or 1. For primary lateral fields, 95\% of apertures were clinically acceptable. For left and right boost lateral fields, 85\% and 90\% of apertures were clinically acceptable, respectively. All the clinically unacceptable apertures were scored as 3, meaning only minor edits were needed to make apertures clinically acceptable.\par
\begin{figure}
\centering
  \includegraphics[scale=.4]{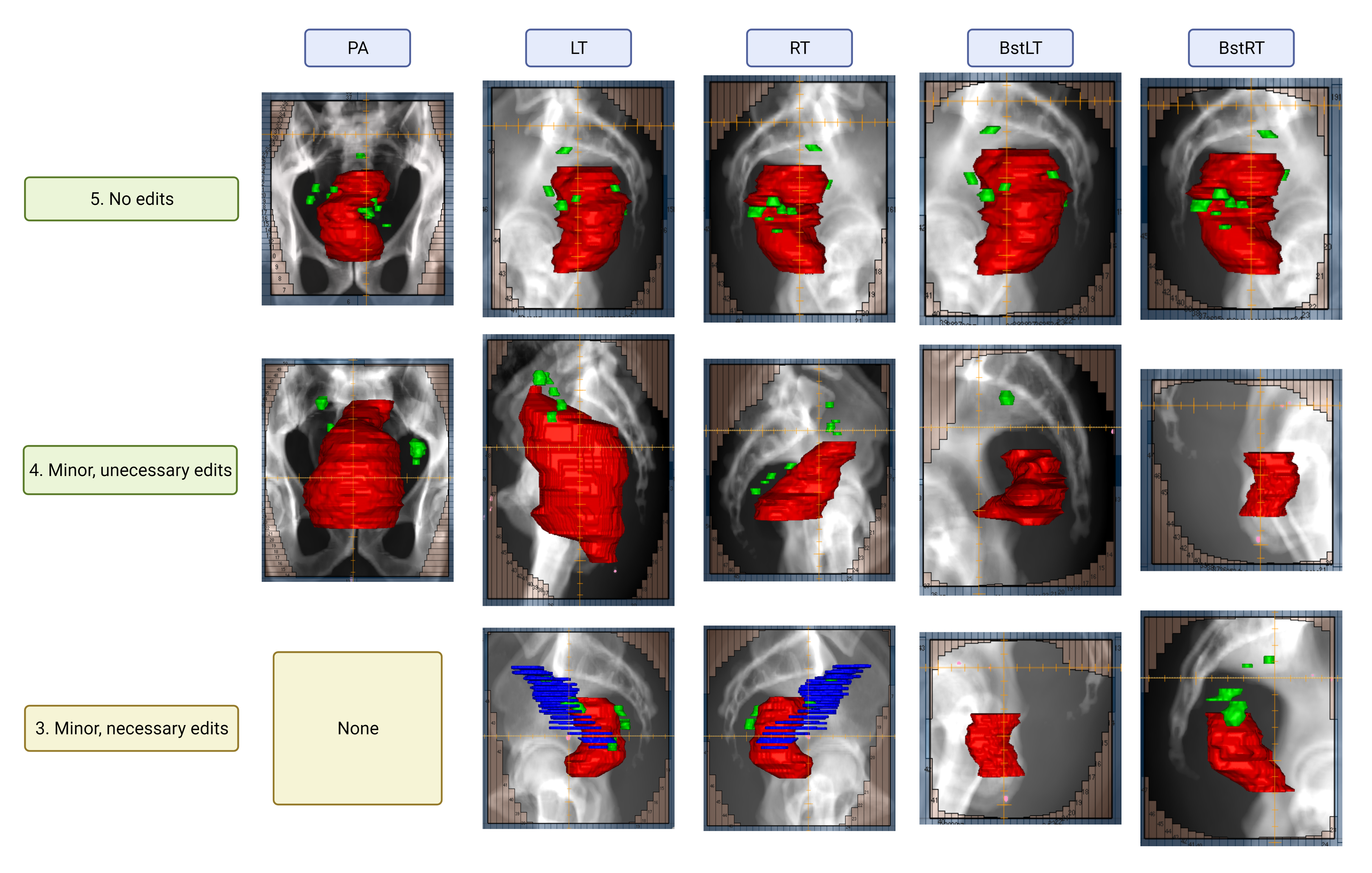}
  \caption{The figure shows examples of field aperture prediction for each field at different scores.Red, green, and blue structures are gross tumor volume (GTV), involved lymph nodes (GTVn), and non-diseased lymph nodes, respectively. The non-diseased lymph nodes were included due to naming error and caused the predicted aperture to have larger anterior opening than desired.}
  \label{fig:fig5}
\end{figure}
Figure \ref{fig:fig5} shows examples of field aperture prediction after MLC fitting for each field at different scores. All primary lateral fields were rated as acceptable except for one needing minor edits because of the inclusion of noninvolved lymph nodes due to an error in structure naming. Each aperture was scored separately for each patient (a total of five apertures per patient); therefore, sometimes apertures for the same patient did not receive the same score. The most common reason for rating boost fields as unacceptable was excessive coverage in the anterior direction.

\subsection{Field-in-field}

\begin{table}
	\caption{The median, mean, and standard deviation of various parameters assessed for each setting tested}
	\centering
\begin{tabular}{llllllllll}
\toprule
                         & \multicolumn{3}{l}{Score} & \multicolumn{3}{l}{Number   of subfields} & \multicolumn{3}{l}{Time per   each field (min:s)} \\
\midrule                        
Setting                  & Median   & Mean   & $\sigma$     & Median         & Mean        & $\sigma$          & Median           & Mean           & $\sigma$             \\
\midrule
45-degree   wedge, 107\% & 5.0      & 4.4    & 0.9   & 4.0            & 4.1         & 1.5        & 13:23            & 13:52          & 5:45          \\
60-degree   wedge, 107\% & 5.0      & 4.5    & 0.9   & 3.0            & 2.9         & 1.3        & 13:47            & 15:14          & 6:21          \\
No   wedge, 107\%        & 3.5      & 3.6    & 0.7   & 6.0            & 5.9         & 0.3        & 10:48            & 12:05          & 3:05          \\
45-degree   wedge, 106\% & 4.0      & 4.2    & 0.8   & 4.0            & 3.9         & 1.5        & 9:06             & 10:23          & 4:13          \\
Average                  & 4.0      & 4.2    & 0.9   & 4.0            & 4.2         & 1.6        & 11:03            & 12:48          & 5:11     \\    
\bottomrule
\end{tabular}
	\label{tab:table5}
\end{table}

\begin{table}
	\caption{The number of plans per score based on physician review and the percentage clinical acceptability for each configuration with different wedges and hotspot percentages}
	\centering
\begin{tabular}{llllllllll}
\toprule
                                          & \multicolumn{5}{l}{Counts per score} & \multicolumn{2}{l}{Acceptability percentage} &       &      \\
\midrule                                         
Setting                                   & 5     & 4     & 3     & 2     & 1    & Acceptable           & Unacceptable          & Mean  & $\sigma$    \\
\midrule
45-degree wedge, 107\% hotspot percentage & 12    & 5     & 2     & 1     & 0    & 85\%                 & 15\%                  & 14:06 & 5:52 \\
60-degree wedge, 107\% hotspot percentage & 13    & 4     & 2     & 1     & 0    & 85\%                 & 15\%                  & 15:14 & 6:21 \\
No wedge, 107\% hotspot percentage        & 2     & 8     & 10    & 0     & 0    & 50\%                 & 50\%                  & 12:05 & 3:05 \\
45-degree wedge, 106\% hotspot percentage & 8     & 9     & 2     & 1     & 0    & 85\%                 & 15\%                  & 10:23 & 4:13 \\
\bottomrule
\end{tabular}
	\label{tab:table6}
\end{table}
\begin{figure}
     \centering
     \includegraphics[scale=1]{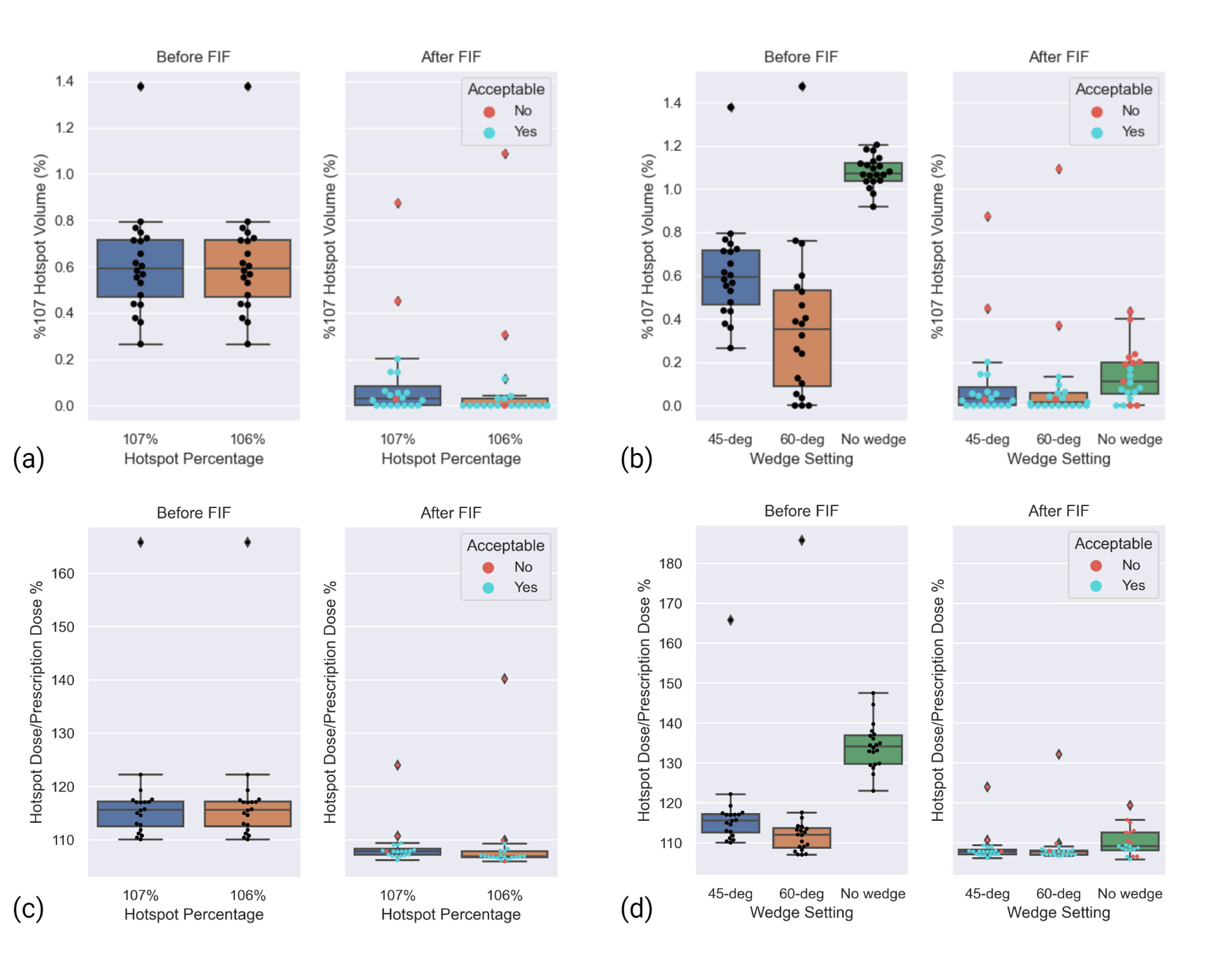}
  \caption{The figure shows the boxplots of plans before and after field-in-field (FIF) for each configuration; (a) Volume exceeding 107\% of Rx as a percentage volume of pseudo-region of interest (pROI) for different hotspot definitions, (b) Volume exceeding 107\% of Rx as a percentage volume of pROI for different wedge settings, (c) For percentage hotspot dose of plans for different hotspot definitions, (d) For percentage hotspot dose of plans for different wedge settings. The plans scored as acceptable ($\geq$ 4) are marked as blue and plans scored as unacceptable ($\leq$ 3) are marked as red.}
\label{fig:fig6}
\end{figure}

\begin{figure}
\centering
  \includegraphics[scale=.7]{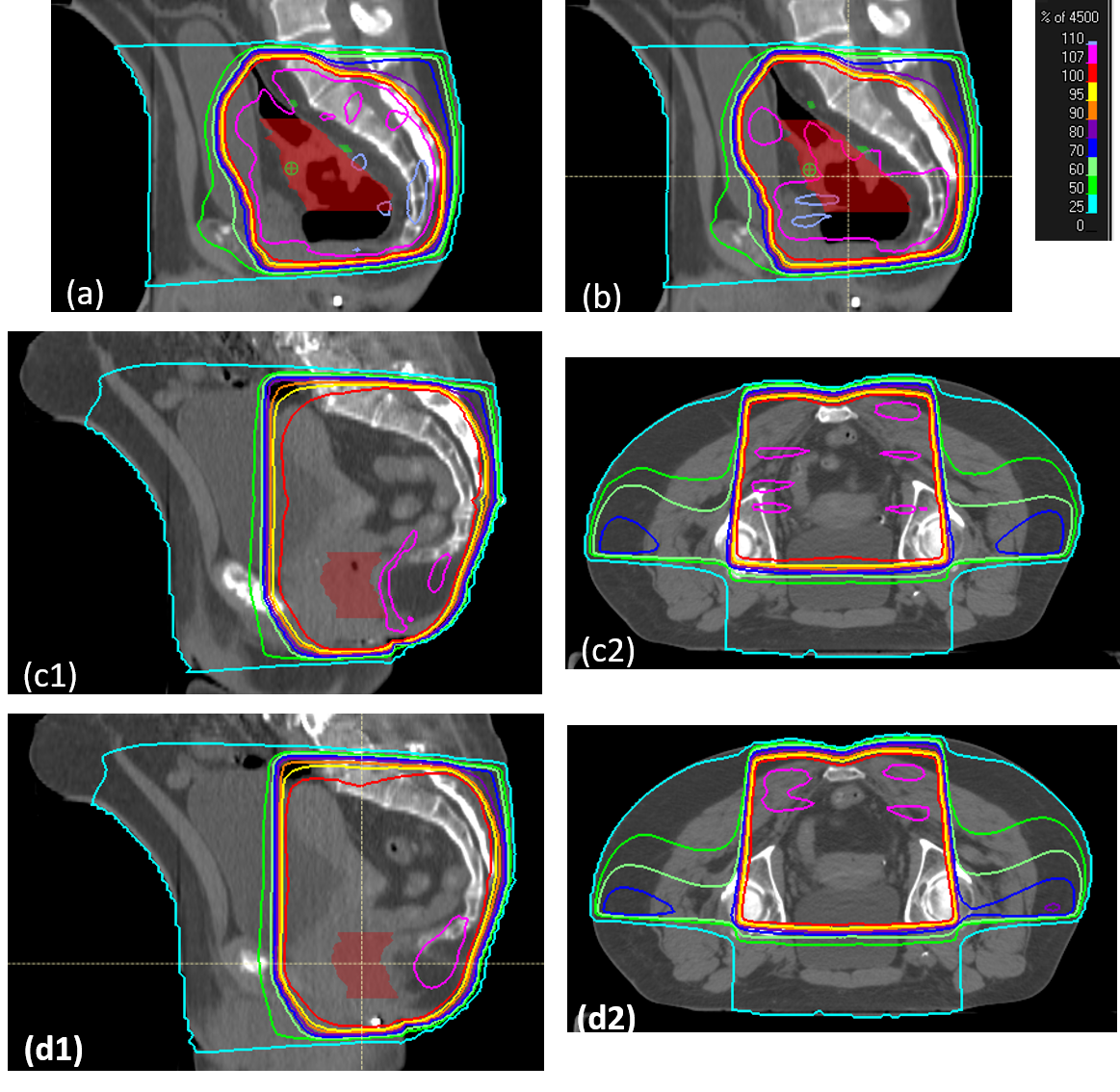}
  \caption{Plans with scores from 2 to 5: Part (a) scores 2 because of large 107\% volume almost covering the entire planning target volume (PTV) area, part (b) scores 3 because the 107\% hotspot was still large, but the plan could be improved by adding an additional subfield, part (c) scores 4 because there was too much 107\% volume in the bowel region. Manually lowering the normalization would allow a score of 5. Parts (c1 and c2) are sagittal and transverse views of the same plan. Part (d) scores 5 with minimal hotspot located in the muscle region. Parts (d1 and d2) are sagittal and transverse views of the same plan.}
  \label{fig:fig7}
\end{figure}
In all the plans made by FIF, the average final hotspot dose percentage was reduced from 121\% ($\sigma$ = 14\%) to 109\% ($\sigma$ = 5\%). The hotspot dose percentage represents the maximum dose percentage in a plan formed by a fully connected volume with at least 8 cm3 in size. In other words, the dose percentage indicated by the hottest 8 cm3. Figure 6a,b shows the V107 as a percentage of pROI. As shown in Figure 6c,d, for plans with 45-degree wedge and 107\% hotspot definition, FIF reduced average hotspot dose percentage from 117.6\% ($\sigma$ = 11.8\%) to 108.7\% ($\sigma$ = 3.8\%). For plans with 60-degree wedge and 107\% hotspot definition, FIF reduced average hotspot dose percentage from 115.0\% ($\sigma$ = 16.9\%) to 108.9\% ($\sigma$ = 5.5\%). For plans with no wedge and 107\% hotspot definition, FIF reduced average hotspot dose percentage from 134.2\% ($\sigma$ = 5.7\%) to 110.2\% ($\sigma$ = 3.5\%).For plans with 45-degree wedge and 106\% hotspot definition, FIF reduced average hotspot dose percentage from 117.6\% ($\sigma$ = 11.8\%) to 108.8\% ($\sigma$ = 7.4\%).\par
On average, 4.2 ($\sigma$ = 1.6) subfields were added to each plan, and it took $\sim$13 ($\sigma$ = 5 min) to add each field (Table \ref{tab:table5}). In the end, the physician review scores for all plans were 4.2 ($\sigma$ = 0.9) on average (Table \ref{tab:table6}).The use of a wedge increased the clinical acceptability from 50\% to 85\% (Table \ref{tab:table6}). Decreasing the definition of hotspot did not influence the overall clinical acceptability, but fewer plans were scored as 5 than 4 because of the tighter coverage as a result (Table \ref{tab:table6}).\par 
Plans with 60- and 45-degree wedges both had an 85\% acceptance rate. However, plans with 60-degree wedges received the highest score of 4.5 on average, and the fewest subfields were added (Table \ref{tab:table5}). Therefore, the 60-degree configuration was used for end-to-end evaluation. Among the plans with varying wedge angles (including 0-degree), at least one plan was scored as acceptable for each patient. For the two patients with whom both wedged plans were found unacceptable, plans with no wedge were acceptable. This means non-wedged plans were superior to wedged plans for those two patients.\par
Regardless of the settings, for all the acceptable plans,the 107\% hotspot volume after FIF comprised less than 20\% volume of pROI (Figure \ref{fig:fig6}a,b) or 500 $cm^3$. Figure 6c,d shows that none of the plans with more than 110\% hotspot dose percentage after FIF was considered clinically acceptable. For the non-wedged plans, the average number of subfields was 5.9, and the average maximum hotspot percentage was 110.2\%. This indicated that the hotspots on some of the non-wedged plans can be further reduced by increasing the maximum number of subfields allowed. For all the plans evaluated, the plans scored as 4 could be rated as 5 after reduction of the normalization to the 100\% isodose line covering 98\% of PTV instead of 99\% of PTV. Examples of plans with different scores and the reason for the score are shown in Figure \ref{fig:fig7}.

\subsection{End-to-end}
Because the left and right fields were predicted separately using respective models, the aperture prediction results did not mirror each other. For end-to-end testing, while adding fields into the TPS, we imported only the right fields for both primary and boost then created an opposing left field based on the imported right field. This guaranteed the lateral fields were mirrored images of each other. \par
\begin{figure}
     \centering
     \begin{subfigure}{0.45\textwidth}
         \centering
         \includegraphics[scale=.45]{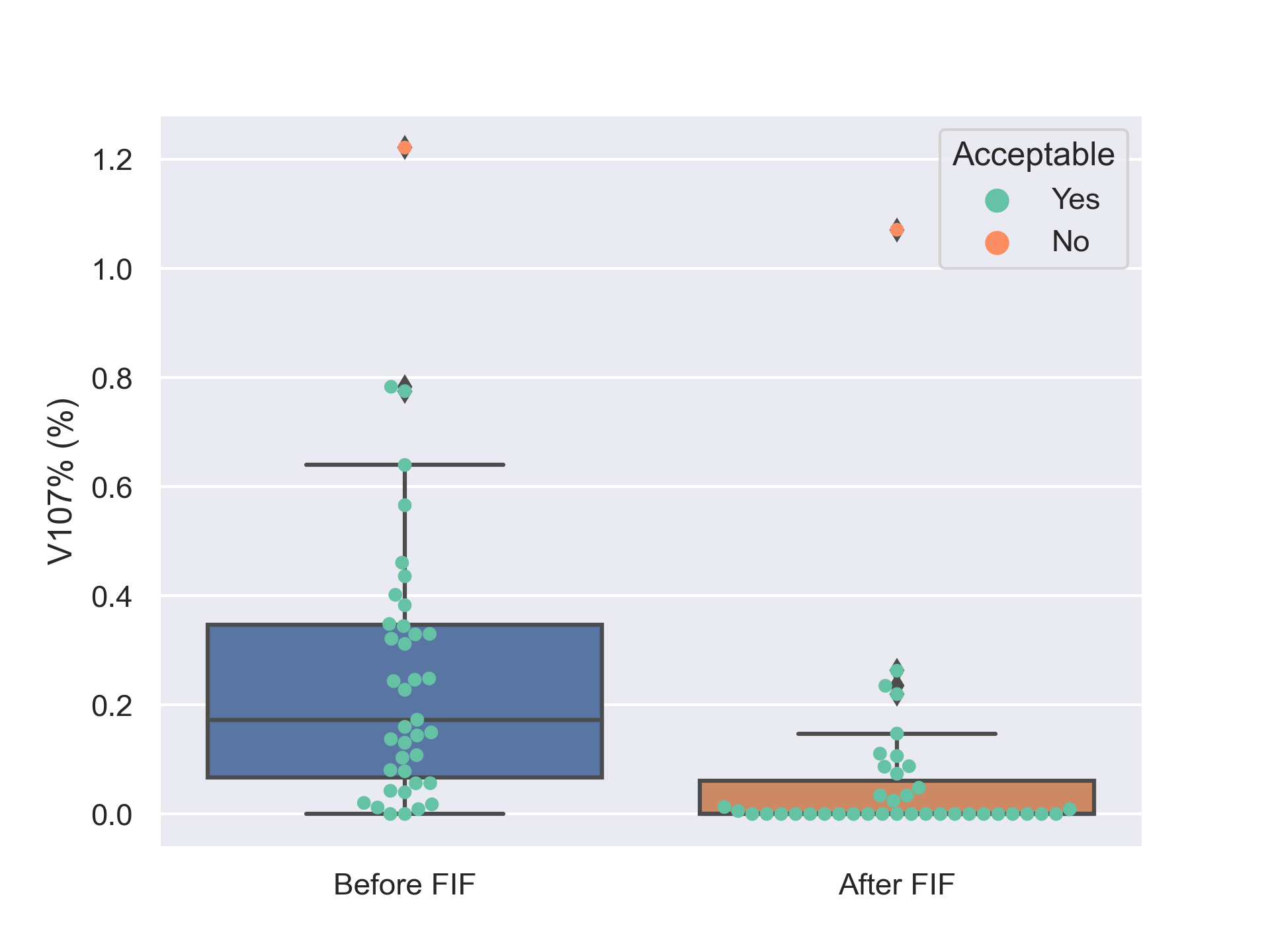}
         \label{fig:fig8a}
     \end{subfigure}
     \begin{subfigure}{0.45\textwidth}
         \centering
         \includegraphics[scale=.45]{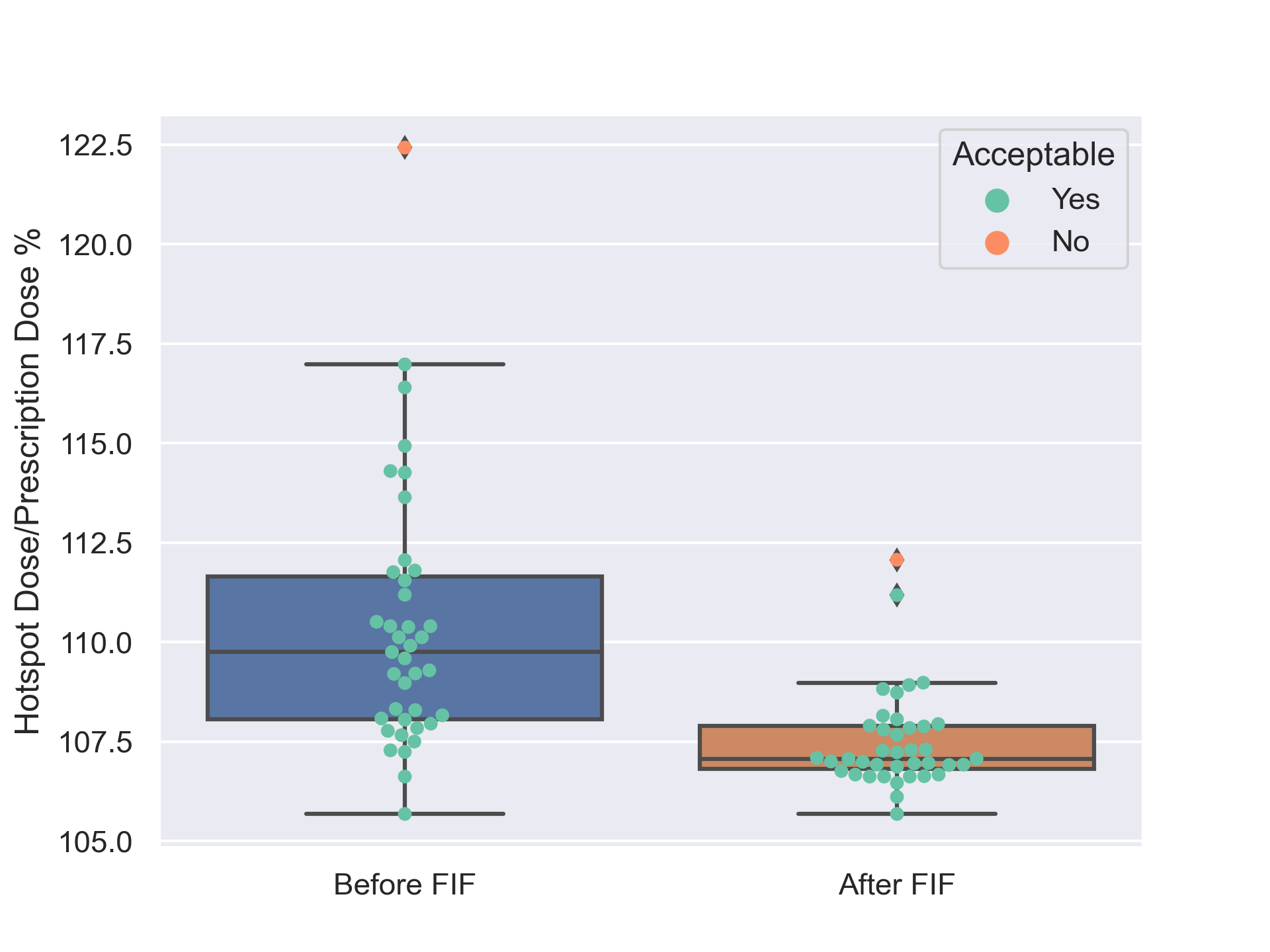}
         \label{fig:fig8b}
     \end{subfigure}
        \caption{The figure shows (left) the volume exceeding 107\% Rx as a percentage of pseudo-region of interest (pROI), that is, V107\% (\%) and (right) the percentage hotspot of primary fields before and after field-in-field (FIF) in end-to-end testing.}
        \label{fig:fig8}
\end{figure}

Figure \ref{fig:fig8} shows the percentage volume of 107\% hotspot, that is V107\% (\%), with respect to pROI and the hotspot dose percentage before and after FIF for primary fields.Table 7 shows that 38 out of 39 plans tested were clinically acceptable. One plan was unacceptable because of the limited aperture margin to the body in the posterior of the sacrum preventing achieving the desired normalization coverage.However, the FIF algorithm produced an acceptable plan for the patient with 100\% Rx covering 98\% of pROI instead of 99\%.
\begin{table}
	\caption{The table shows the result of physician review for end-to-end testing including the number of plans per score based on physician review and the percentage of clinical acceptability.}
	\centering
\begin{tabular}{llllllll}
\toprule
End-to-end testing                        & \multicolumn{5}{l}{Counts per score} & \multicolumn{2}{l}{Acceptability percentage} \\
\midrule
Setting                                   & 5      & 4      & 3    & 2    & 1    & Acceptable (\%)           & Unacceptable (\%)          \\
\midrule
60-degree wedge, 107\% hotspot percentage & 23     & 15     & 1    & 0    & 0    & 97.4               & 2.6                \\
\bottomrule                                                         
\end{tabular}
	\label{tab:table7}
\end{table}

\section{Discussion}
The results of the field aperture prediction showed that given accurate contours of the GTV, GTVn, and DRR, a DL model can produce field apertures that are clinically acceptable.Post-processing was not applied to the predicted field apertures because MLC fitting smoothed out the small unevenness in the shapes. All the aperture models achieved Dice similarity scores higher than 0.9. However, high Dice scores are not closely linked with clinical acceptability because the actual position of the difference is not reflected in the Dice score. The HD and MSD values indicated that differences between predicted aperture and clinical apertures were in the order of less than 1.9 cm in the maximum distances and 0.6 cm in average distances. The differences in distances tended to be due to the large variations in the anterior aspect of aperture shapes. The values also showed that there were more variations in boost apertures than primary apertures in practice. However, large HD and MSD values to clinical apertures also did not necessarily indicate low clinical acceptability as the results of physician review for apertures showed that DL-generated apertures were consistent, and the clinical acceptability was high. Therefore, physician reviews are critical in evaluating clinical acceptability.\par
Models for lateral fields and boost fields can produce noticeably different aperture shapes for the opposing beam angles. This effect may be exacerbated by metal implants, seen in one of the test patients. Further studies can be performed to identify the large differences in aperture shapes as quality assurance to flag uncertain predictions. As a part of an automatic routine, it is also important to establish standardized names for nodes and GTV such that non-diseased volumes are not added to the prediction, resulting in aperture shapes that need editing.\par
For cases in which attending physicians have decided the apertures following the guidelines are not appropriate, the aperture shapes may be modified before the FIF algorithm is used. If the guidelines change, then specific data adhering to the new guidelines will need to be curated and the model for aperture prediction updated and reviewed thoroughly.\par
There are different approaches to manually creating FIF plans.\cite{ercan_dosimetric_2010,lo_intensity_2000,lee_performance_2008} The definition of hotspots as a percentage of Rx depends on the treatment sites and the clinical practices and could range from 102\% to 107\% of Rx.\cite{ercan_dosimetric_2010,lo_intensity_2000,lee_performance_2008} Our method allows users to configure definitions of hotspots. At each iteration, subfields can be made based on a fixed percentage of a hotspot, for example, reducing a 107\% hotspot iteratively till it is nonexistent, or the percentage of the hotspot can decrease at each iteration, for example, subsequently creating subfields based on 115\%, 110\%, and 105\% Rx. The program is capable of both using a fixed percentage of hotspots and dynamically reducing the percentages at each iteration, though for rectal, we found that having a fixed hotspot definition at each iteration was sufficient and could produce consistent results with high clinical acceptability.\par
The results demonstrated that FIF plans can be added in a forward-planning manner without human intervention for rectal cancer radiotherapy. For most of the patients, plans using 60-degree wedges are clinically acceptable. Overcompensation was the primary reason that the plans with wedges were unacceptable. Overcompensation could be caused by patients having a large un-overridden volume of gas, relatively smaller body curvature in the AP direction on the lateral sides, and either a tight aperture margin to the GTV in the anterior direction or aperture being too close to the skin in the posterior direction. These would result in extremely hot plans in order to achieve the desired normalization shown as outliers in Figure \ref{fig:fig6}. Overcompensation could be remedied by using a non-wedge configuration or reducing the normalization of plans before applying FIF.\par
To our knowledge, this is the first automated treatment planning tool designed for rectal 3DCRT. Previous works have automated the 3DCRT planning process for breast cancers with FIF.\cite{kim_automated_2018,kisling_automated_2019} Though the results were not directly comparable, the time it takes for automating a plan for breast cancer with FIF is generally faster than for rectal cancers 3DCRT with FIF. This might be caused by the difference in field geometries in breast and rectal cancer treatments and the use of beam weight optimization in our study. The end-to-end algorithm was designed to create plans in parallel, allowing multiple plans to be created at once. The entire process does not need user interventions,allowing clinical professionals to focus on more challenging cases. For creating a single plan, most of the overhead time was spent on optimization for beam weights. Further studies can seek to reduce the optimization time by using different optimization solvers. Previous studies have compared manual FIF plans with automatically produced FIF plans for breast cancers.\cite{kim_automated_2018,kisling_automated_2019} In this study, we have focused our efforts on validating the end-to-end solution based on physician evaluations. Future studies may introduce a direct comparison between the automatically generated FIF plans versus the manual plans for rectal cancers.\par
Furthermore, different clinics can have different clinical requirements for dose homogeneity and target coverage. In this study, we have demonstrated the effectiveness of the methods for our clinical standard practice. For example, a wedge of 15-degree was used for boost fields, as this is standard practice at our institution, but other wedges could also be used.To translate the entire workflow to other clinics, similar studies of quantitative and qualitative evaluations of the outputs need to be conducted using local data and with practicing physicians. For example, specific data need to be curated for field aperture prediction, and a different configuration may need to be used for the FIF program. To ensure clinical acceptability, additional physicians familiar with local practices should assess plans prior to deployment. Furthermore, the FIF algorithm may have the potential to be translated to radiotherapy for other body sites, such as whole-brain irradiation. Finally, given that we have tested field aperture prediction and the FIF program on 20 patients and performed end-to-end testing on 39 patients, future studies may increase the number of patients tested to examine the robustness of the method in larger and different populations. Though beyond the scope of this study,any clinical integration of automated tools should undergo extensive stress testing and regular performance checks to ensure safety and consistency.

\section{Conclusion}
The study proposed, implemented, and tested an automated solution to create apertures and add subfields for rectum 3DCRT plans. The study demonstrated the effectiveness of the automated solution by evaluating the creation of aperture and FIF plans both independently and as a whole. The program requires no human interaction and is TPS independent. The study demonstrates that the FIF technique can be automated and subfields can be added in a forward-planning method to produce high-quality clinical plans.

\section*{Acknowledgments}
The authors would like to thank Sarah Bronson, ELS, of the Research Medical Library at The University of Texas MD Anderson Cancer Center for her editing of the manuscript. The authors would also like to thank the funding support provided by the MD Anderson Cancer Center Institutional Research Grant.

\section*{Conflict of interest}
The authors declare that there is no conflict of interest that could be perceived as prejudicing the impartiality of the research reported.

\bibliographystyle{unsrt}
\bibliography{references}  






\end{document}